\documentclass[11pt,twoside]{article}
\usepackage{FUSEConf2004}
\usepackage{natbib}

\usepackage{lscape}
\usepackage{epsfig}

\begin{document}
\title{Deuterium at High Redshifts: Recent Advances and Open Issues}
\author{Max Pettini}
\affil{Institute of Astronomy, Madingley Road, Cambridge, CB3 0HA, England}

\begin{abstract}
Among the light elements created in the Big Bang, deuterium
is one of the most difficult to detect but is also the one
whose abundance depends most sensitively on the density of baryons. 
Thus, although we still have only a few positive identifications
of D at high redshifts---when the D/H ratio was close to its primordial 
value---they give us the most reliable determination of the baryon density,
in excellent agreement with measures obtained from
entirely different probes, such as the anisotropy of the cosmic
microwave background temperature and the average absorption
of the UV light of quasars by the intergalactic medium.
In this review, I shall relate observations of D/H in distant 
gas clouds to the large body of data on the local abundance 
of D obtained in the last few years with {\it FUSE\/}. I shall also
discuss some of the outstanding problems in light element
abundances and consider future prospects for advances 
in this area.
\end{abstract}

\section{Introduction and Historical Background}

The measurement of the interstellar abundance of deuterium was
one of the main science drivers of the \emph{FUSE} mission right from 
its inception. Five years on, this promise has been amply fulfilled,
as demonstrated by the numerous talks and posters at this
meeting devoted to \emph{FUSE} results on D/H.

The importance of deuterium stems from the fact that, among
the light elements created in Big Bang nucleosynthesis (BBNS), 
it is the one whose primordial abundance responds most sensitively
to cosmological density of baryons, $\Omega_{\rm b}$.
While $^4$He is the most abundant, because it soaks up essentially 
all the available neutrons,
this property also makes it a rather insensitive 
`baryometer'. The quantity $\Omega_{\rm b}$, or more
precisely the baryon to photon ratio $\eta$,
only affects $Y_{\rm p}$ (the primordial mass fraction in  $^4$He) 
by determining the time delay before BBNS can set in---and 
thus the time available for neutrons to decay---in the first 
few minutes of the universe history.
The more fragile deuterium, on the other hand, is easily destroyed
by two-body reactions with protons, neutrons and other nuclei 
so that its abundance relative to hydrogen 
when BBNS is over, (D/H)$_0$ or D$_0$ for short,
shows a steep, inverse, dependence on $\Omega_{\rm b}$.
$^7$Li is less useful than D in this respect because it is far
less abundant, by about five orders of magnitude, and its dependence
on $\Omega_{\rm b}$ is double-valued because it can be synthesised via different
nuclear reactions in the high and low baryon density regimes.

The detection of interstellar D was among the first  
discoveries made by \emph{FUSE}'s predecessor, the \emph{Copernicus}
satellite. Rogerson \& York (1973)
resolved the isotope shift in the Ly$\gamma$ line seen towards
the bright B1 III star $\beta$~Centauri, and deduced
$N$(\ion{D}{1})/$N$(\ion{H}{1})$\,= (1.4 \pm 0.2) \times 10^{-5}$.
Three decades later, the mean of 21 measurements
of D/H in the `Local Bubble' (the nearby region of the Milky Way disk)
is in excellent agreement with \emph{Copernicus}' first detection:
$\langle$D/H$\rangle = (1.56 \pm 0.04) \times 10^{-5}$
(Wood et al. 2004).\footnote{An interesting observation is that
the distance to $\beta$~Cen has `doubled' since 1973. 
The \emph{Hipparcos} parallax to this star implies a distance
$d = 161\,$pc, while the parallactic distance available to
Rogerson \& York (1973) was $d = 81$\,pc. 
This is a clear demonstration
that it is easier for astronomers to measure chemical abundances
than distances, even to the brightest stars.}

An important property of deuterium is that it is only destroyed whenever
interstellar gas is cycled through stars (a process termed astration), 
so that its abundance relative to H steadily decreases with
the progress of galactic chemical evolution. Analytically,
this reduces to a simple expression for the time evolution
of D:
\begin{equation}
{\rm (D/D_0)}~ = ~f^{(1/\alpha~ -1)}~ = ~e^{-Z(1/\alpha~ -1)}
\end{equation}
where $f$ is the gas fraction, $Z$ the metallicity
(in units of the yield of a primary element such as oxygen)
and $\alpha$ is the mass fraction 
which is locked up in long-lived stars and stellar remnants
whenever a quantity $M$ of interstellar matter is turned into stars
(Ostriker \& Tinsley 1975; Pagel 1997). 
Equation (1) is valid in the simplest case
of a `closed-box' model of chemical evolution. More
realistic models which include inflow and/or outflow generally
result in lower reductions of the primordial D/H 
as a function of either $f$ or $Z$ (Edmunds 1994).
We cannot measure the lock-up fraction $\alpha$ directly,
but only deduce it theoretically by assuming a distribution
of stellar masses (the IMF) and guessing at what fraction
of its initial mass each star returns to the interstellar medium (ISM).

In principle, the degree of astration suffered by D in the Milky Way,
where $f = 0.1 - 0.2$, could be anywhere between 20\%
and 90\%, depending on the uncertain value of $\alpha$.
Consequently, Rogerson and York could only
use their measurement of D/H in the ISM today to place a lower
limit on (D/H)$_0$ and a corresponding 
upper limit 
$\Omega_{\rm b} \leq 0.0675$ (for $h = 0.7$ where, as usual,
$h$ is the Hubble constant in units of 100\,km~s$^{-1}$~Mpc$^{-1}$.
In this article I shall use $h = 0.7$ throughout and 
dispose of the $h^2$ term in the value of $\Omega_{\rm b}$).
Note that the above upper limit on $\Omega_{\rm b}$, 
even without any correction for D astration, 
implies that most of the matter in the universe
is non-baryonic.

\section{Deuterium at High Redshifts}

In parallel with the rapid advances in our knowledge of the Galactic
interstellar medium made possible by the success of the
\emph{Copernicus} mission, the mid-1970s saw the burgeoning of
QSO absorption line spectroscopy which extended similar lines
of enquiry to the gas in galaxies and the intergalactic medium
at much earlier times. Thanks to expansion of the universe, which
redshifts the same ultraviolet transitions identified by
\emph{Copernicus} into the optical region, the interstellar
media of distant galaxies which happen to lie in front of 
quasars could begin to be probed systematically,
by capitalising on the light-gathering power 
of ground-based telescopes, as well as the efficiency and linearity 
of recently developed digital detectors.
Adams (1976) was the first to point out that such observations
were also likely to bring us closer to determining the primordial
abundance of deuterium, since gas less chemically evolved than the
local ISM should be more common at high redshift.

In the mid-1970s the practical difficulties of detecting D in QSO
absorption line systems were daunting. The combined requirements 
of high spectral resolution (the isotope shift in the Lyman series
amounts to $-82$\,km~s$^{-1}$), high sensitivity (even the brightest
high redshift QSOs are more than a million times fainter than
$\beta$~Cen), and wide wavelength coverage (so as to record 
several Lyman lines at once) could not be met for another twenty 
years, until the advent of echelle spectrographs on 8-10\,m class
telescopes. Even with these technological advances, however, 
isolating D at high $z$ remains an intrinsically difficult observation
due to (a) the high density of H absorption lines in the Ly$\alpha$ forest
(see Figure 1) and (b) the paucity of QSO absorption systems with
a sufficiently simple velocity structure to resolve cleanly \ion{D}{1} 
absorption blueshifted by 82\,km~s$^{-1}$ from its corresponding \ion{H}{1}. 
All but a few percent of either
Lyman limit systems (LLS), absorbers with column densities 
$N$(\ion{H}{1})$\, \ga 3 \times 10^{17}$\,cm$^{-2}$,
or damped Ly$\alpha$ systems (DLAs) with 
$N$(\ion{H}{1})$\, \geq 2 \times 10^{20}$~cm$^{-2}$,
exhibit multiple velocity components spanning more than 100\,km$^{-1}$.
And yet these two classes of QSO absorbers,
at the high end of the power-law distribution of \ion{H}{1}
column densities, are the ones which can be realistically targeted in 
D searches, given that the primordial abundance of D
is likely to be of the order of a few times $10^{-5}$.

%
% FIGURE 1
%
\begin{figure}[!ht]
%\vspace*{-2cm}
\hspace*{0.25cm}
\plotfiddle{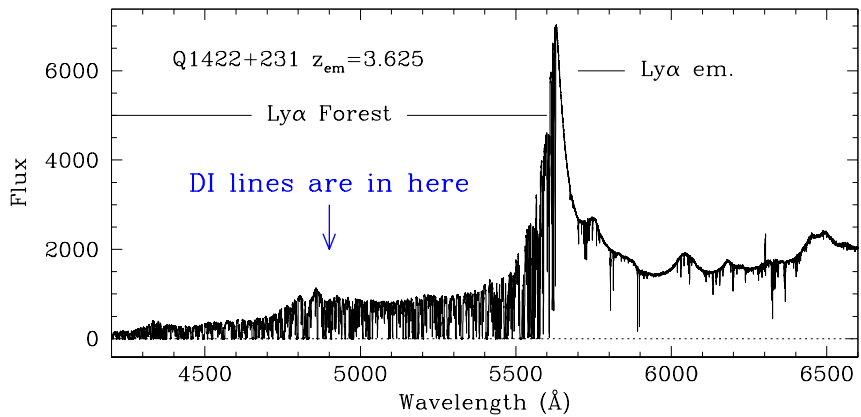}{5.5cm}{270}{50}{50}{-187}{230}
\vspace{0.15cm}
\caption{High resolution spectrum of a typical high redshift QSO (reproduced from
the work of Ellison et al. 2000). At redshifts $z > 2$ it becomes progressively more
difficult to distinguish \ion{D}{1} absorption from the hundreds of absorption lines which make
up the Ly$\alpha$ forest.
}
\label{}
\end{figure}

These obstacles partly explain why, ten years after the first
detection of D in a high redshift QSO absorber by
Tytler et al. (1995), the number of measurements generally 
regarded as reliable is disappointingly low (see Figure 2).
Averaging the five published values of (D/H) 
obtained from D lines which
are at least partially resolved (Kirkman et al. 2003),
Steigman (2004) deduced 
$\langle {\rm (D/H)_0} \rangle = (2.6 \pm 0.4) \times 10^{-5}$
(the error quoted is $\sigma/\sqrt{5}$).
This is likely to be the primordial value, since all
five absorption systems originate in gas of low
metallicity---1/30 of solar or less (see Figure 2)---where
according to equation (1) the astration of D should have been insignificant.
Using the scaling of (D/H)$_0$ with $\eta$ and $\Omega_{\rm b}$ 
given by Burles, Nollett, \& Turner (2001),
${\rm (D/H)_0} = (2.6 \pm 0.4) \times 10^{-5}$
implies $\Omega_{\rm b} = (0.044 \pm 0.004)$.

%
% FIGURE 2
%
\begin{figure}[!ht]
\vspace*{-0.5cm}
\hspace{1.5cm}
\plotfiddle{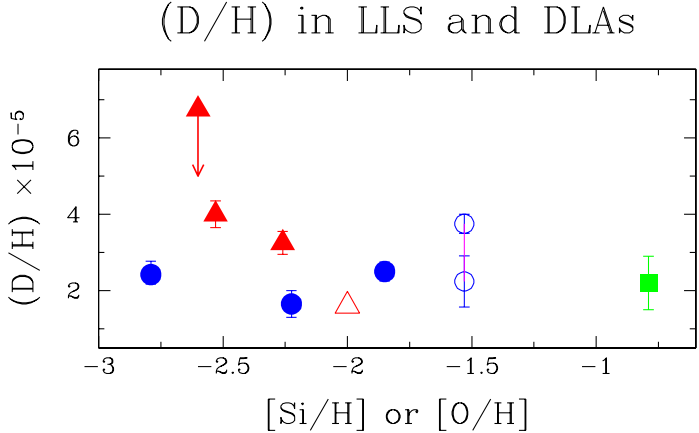}{5.5cm}{270}{60}{60}{-225}{250}
\vspace{0.35cm}
\caption{
Compilation of D/H measurements in QSO absorbers (as of August 2004).
Triangles denote measurements in Lyman limit systems, while circles are for damped
Ly$\alpha$ systems. Filled symbols are used for cases where the \ion{D}{1} absorption
is at least partially resolved (O'Meara et al. 2001; Kirkman et al. 2003; Pettini \& Bowen 2001).
Open symbols indicate absorption systems with more complex velocity structure which renders
the determination of D/H less straightforward (open triangle: Crighton et al. 2004; open circles:
D'Odorico, Dessauges-Zavadsky, \& Molaro 2001; Levshakov et al. 2002). 
The two open circles refer to two analyses of the
same absorber, at $z_{\rm abs} = 3.02486$ in Q0347$-$383, and illustrate the uncertainty in D/H
resulting from different interpretations of the multi-component character of the 
absorption lines. For this reason, the currently favoured value of (D/H)$_0$ is generally
taken to be the weighted mean of the five detections indicated by the filled triangles and circles
(see text). The filled square shows, for comparison, the \emph{FUSE} measurement by Sembach et 
al. (2004) of D/H in Complex C, a high Galactic latitude concentration of \ion{H}{1}
which is thought to be relatively unprocessed gas being accreted by the Milky Way.
Plotted on the $x$-axis is the abundance of either Si or O in the absorber in the customary
notation whereby [Si/H]\,$ = \log$(Si/H)$_{\rm abs} - \log$(Si/H)$_{\sun}$.
All high redshift QSO absorbers where D/H has been measured are chemically unevolved
systems, with metallicities of less than $\sim 1/30$ of solar.
}
\label{}
\end{figure}

\section{Other Measures of $\Omega_{\rm b}$}
Recent years have seen spectacular advances in the 
determination of a number of fundamental cosmological
parameters, including $\Omega_{\rm b}$.
In particular, the {\it WMAP\/} satellite and other experiments
have now mapped with high precision the temperature anisotropies 
in the cosmic microwave background (CMB)
over a range of angular scales which includes the first few
peaks in the power spectrum.
Their relative amplitudes vary with $\Omega_{\rm b}$, as can be seen from
Figure 3. The physical reason for this is that the baryon density 
determines the inertia in the photon-baryon fluid. 
Higher values of $\Omega_{\rm b}$ result in 
deeper compressions and less pronounced rarefaction; thus the
compressional peaks (the odd-numbered ones) are hotter, while
the rarefaction peaks (even numbered) are cooler (Page et al. 2003).
The best fit to the CMB temperature angular power spectrum
is obtained for $\Omega_{\rm b} = (0.045 \pm 0.002)$ (Spergel et al. 2003)
in near-perfect agreement with the value implied by the primordial
abundance of deuterium. Sometimes we take our achievements for granted.
The fact that we can measure the cosmological density of ordinary matter
in two totally independent ways---one based on a set of nuclear reactions
which took place in the first few minutes in the existence of our universe,
the other on the acoustic oscillations in the mix of photons, dark matter
and baryons which became imprinted on the microwave sky 
some 380\,000 years later---and get the same answer is a spectacular
success of modern observational cosmology 
and gives us confidence in the validity of the entire framework.

%
% FIGURE 3
%
\begin{figure}[!ht]
\vspace{0.25cm}
\hspace{2.75cm}
\plotfiddle{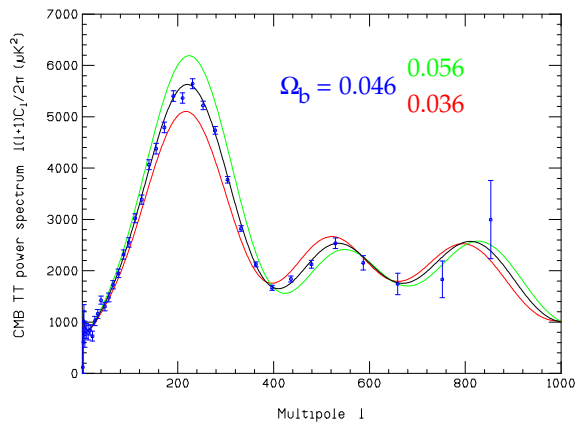}{4.75cm}{0}{38}{38}{-107}{-10}
\vspace{0.15cm}
\caption{The data points are the temperature anisotropies 
in the cosmic microwave background 
measured by the \emph{WMAP} satellite. The variance of the
multipole amplitude is plotted vs. multiple number $\ell$\/ 
(the angular scale on the sky corresponding to multipole $\ell$ 
is $\theta \sim 200\deg /\ell$\/).
The continuous curves show the sensitivity of the power spectrum
to $\Omega_{\rm b}$, while keeping all other relevant cosmological 
parameters fixed. (Figure reproduced from Steigman 2004).
}
\label{}
\end{figure}

The average flux decrement in the Ly$\alpha$ forest, measured by the 
parameter {\it DA\/} first introduced by Oke \& Korycansky in 1982 (see Figure 4),
is also sensitive to the baryon density. The recent comprehensive 
analysis by Tytler et al. (2004) arrived at a best estimate
$DA(z = 1.9) = 0.151 \pm 0.007$ in the interval $1070 < \lambda_0 < 1170$\,\AA, or
$DA(z = 1.9) = 0.12 \pm 0.01$ after correcting for metal lines, LLS and DLAs.
Hydrodynamic simulations can reproduce this value of {\it DA\/} if
$\Omega_{\rm b} = 0.044 \pm 0.002$.

%
% FIGURE 4
%
\begin{figure}[!ht]
%\vspace{0.25cm}
\hspace{1.6cm}
\includegraphics[width=0.725\textwidth]{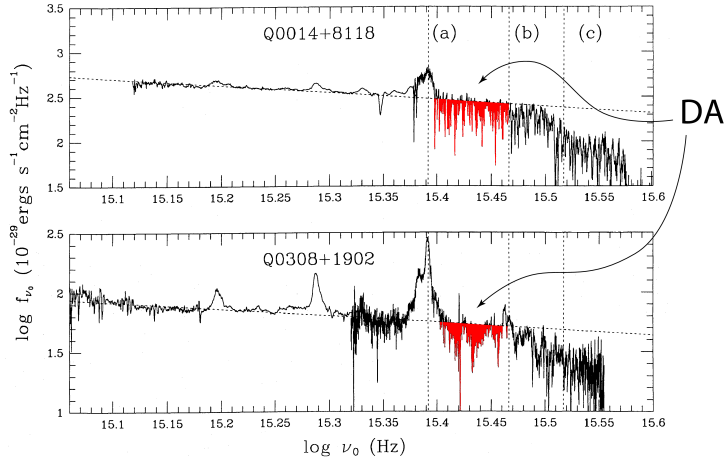}
%\plotfiddle{mp_fig4.eps}{4.75cm}{0}{60}{60}{-175}{-200}
%\vspace{0.7cm}
\caption{The parameter {\it DA\/} measures with a single number
the average opacity of the Ly$\alpha$ forest. Hydrodynamic simulations
of the intergalactic medium show that the value of {\it DA\/} 
depends on the combination of the mean gas density ($\Omega_{\rm b}$),
the density fluctuations (the parameter $\sigma_8$),
and the intensity of the ionising background.
The two QSO spectra shown here are from the survey by 
Steidel \& Sargent (1987).
}
\label{}
\end{figure}

\section{Implications}
The excellent agreement between the three independent 
determinations of $\Omega_{\rm b}$ discussed above could reasonably
be taken as evidence that this 
cosmological parameter is now known to better than 10\%. 
However, in order to be confident that this is indeed the 
case, it is important to carry out as many consistency
checks as possible by considering the consequences of adopting
$\Omega_{\rm b} = 0.044$\,.

\subsection{Is the Degree of Astration of D Plausible?}
Adopting the value 
${\rm D/H} = (1.56 \pm 0.04) \times 10^{-5}$ 
from Wood et al. (2004) as representative of the
D abundance in the local interstellar medium,
and a primordial ${\rm (D/H)_0} = (2.6 \pm 0.4) \times 10^{-5}$
(Steigman 2004) implies that
less than 50\% of the D created in the Big Bang has been destroyed 
through the entire chemical evolution history of the Milky Way Galaxy.
In our simple `closed-box' model where $(D/D_0)~ = ~f^{(1/\alpha~ -1)}$,
$(D_{\rm ISM}/D_0) = 0.60$ in turn implies $\alpha \simeq 0.75 - 0.8$ (for $f = 0.2 - 0.1$).
As pointed out by Edmunds (1994), a value of $\alpha$
in this range is consistent with expectations for a Salpeter IMF,
using a back-of-the-envelope accounting whereby
all stars less massive than the Sun have not returned any baryons
to the interstellar medium over the lifetime of the Galactic disk
and all stars more massive than the Sun
leave behind $1\,M_{\odot}$ remnants at the end of their lives.
More complex chemical evolution models, 
including infall onto the Milky Way disk, can also 
accommodate this degree of astration within the current picture of
the assembly of the Galaxy and its past history
of star formation (e.g. Tosi et al. 1998; Prantzos \& Ishimaru 2001; 
Romano et al. 2003).

Questions remain, however, as to whether the D/H ratio
measured by Wood et al. (2004) in the Local Bubble is 
representative of the true Galactic deuterium abundance---a topic
which was debated extensively at this meeting. 
Doubts are raised by our inability to find a straightforward
explanation for the unexpectedly large scatter in the values
of D/H revealed by {\it FUSE\/} observations of stars beyond
$\sim 100$\,pc from the Sun. Differing interpretations which have been
put forward include selective depletion of D onto grains
(see J. Linsky's and B. Draine's contributions to this volume)
and recent infall of chemically unenriched gas onto the solar
neighbourhood (see G. H\'{e}brard's article). In the former case
it is advocated that the \emph {highest} measures of D/H are the
representative ones, since presumably they suffer the lowest
amount of dust depletion, and the true (gas+dust) value of 
(D/H)$_{\rm ISM}$ could then be as high as $2.3 \times 10^{-5}$.
It remains to be established whether levels of astration as low as 10--15\%
are plausible for the Milky Way.
In the latter case, it is argued that the low measures of D/H 
(and D/O---see H\'{e}brard \& Moos 2003)
over the larger areas of the disk which lie beyond 
the Local Bubble more closely reflect the 
consumption of D over the chemical history of the Galaxy;
in this case (D/H)$_{\rm ISM} \simeq (0.5 - 1) \times 10^{-5}$
and D$_{\rm ISM}$/D$_{0} \simeq 0.2 - 0.4$\,. 
The lower limit of this range may require strong
gas outflows to have taken place during the early 
evolution of the Milky Way, a scenario which now 
seems unlikely (Edmunds 1994; 
Prantzos \& Ishimaru 2001).

Although five years of \emph{FUSE} observations
have provided us with a wealth of new data on the D/H ratio
in the local ISM, it is very frustrating that the full
picture still eludes us. A possible way out of this impasse 
may be to target  
distant stars whose sight-lines are known to exhibit
widely different degrees of interstellar dust depletions.
A correlation (or absence of one)
between the D/H ratio and the gas-phase abundances 
of the most depleted elements,
such as Fe and Ti, may be the clue which we are still
missing.

\subsection{Tension with the Abundances of Other Light Elements}
While the primordial abundance of D agrees with the most likely estimates
of $\Omega_{\rm b}$ from the CMB and the Ly$\alpha$ forest opacity, 
those of $^4$He and $^7$Li apparently do not (see Figure 5).
Perhaps the idea that 
$Y_{\rm p}$ (the primordial mass fraction in  $^4$He)
can be used as a precise baryometer 
is too optimistic. While the fact that approximately
one quarter of the baryons is in the form of $^4$He
is one of the pillars of the standard Big Bang model,
knowing now that $\Omega_{\rm b} = 0.044$ shows just how insensitive
$Y_{\rm p}$ is to the precise value of $\Omega_{\rm b}$.
Once we are on the flat part of the $Y_{\rm p}$ vs. $\Omega_{\rm b}$ 
curve (see Figure 5),
we are vulnerable to subtle systematic errors in the determination
of the helium abundance in \ion{H}{2} regions---and its extrapolation
to zero metallicity---which may be very difficult to overcome
(e.g. Peimbert et al. 2003). Olive \& Skillman (2004) have recently
re-examined the problem and proposed a non-parametric approach whereby
the physical parameters (in particular the nebular electron temperature, 
the optical depth of the emission lines, and underlying stellar absorption)
which enter in the determination of the
helium abundance are derived in a self-consistent way solely
from the relative ratios of He and H emission lines.
Their main conclusion is that previous analyses have largely
underestimated the systematic uncertainties in the determination
of $Y_{\rm p}$ and that even the best data available at present
cannot constrain $Y_{\rm p}$ to better than $0.249 \pm 0.009$.
While this relieves the `tension' with D$_0$, it also raises questions
as to whether we can ever pin down the primordial helium abundance to
the degree of accuracy required to test concordance with 
other light elements.

%
% FIGURE 5
%
\begin{figure}[!ht]
%\vspace{0.25cm}
\hspace{3.00cm}
\plotfiddle{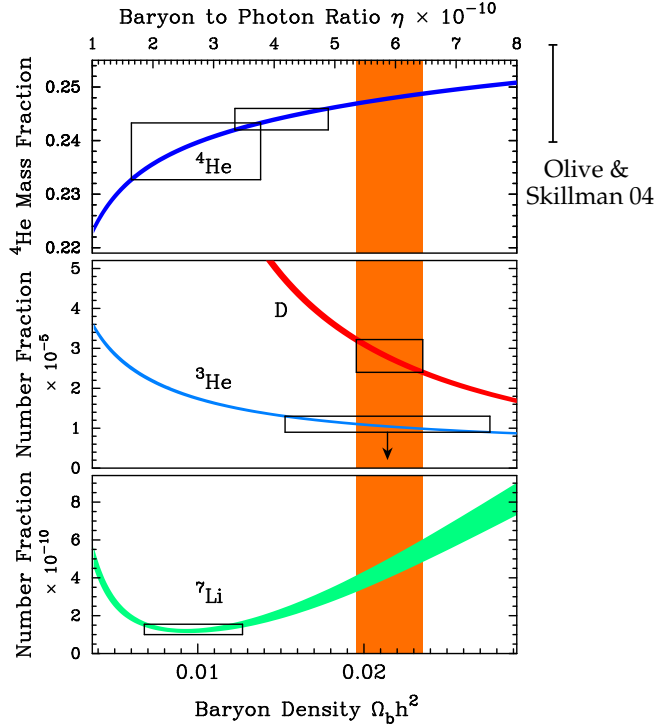}{8.5cm}{0}{37}{37}{-100}{-25}
\vspace{0.85cm}
\caption{Observed and predicted primordial abundances of light
elements created in BBNS (Figure reproduced from Kirkman et al. 2003). 
The vertical band shows the values of $Y_{\rm p}$ and
($^7$Li/H)$_0$ expected for $\Omega_{\rm b} = 0.044$
[the value implied by (D/H)$_0$]. 
Current estimates of $Y_{\rm p}$ by Olive, Steigman, \& Skillman (1997),
shown by the larger box in the top panel, 
and by Izotov \& Thuan (1998) (smaller box) fall short of
the expected value of $Y_{\rm p}$. This may well be due
to underestimated systematic errors, as proposed by Olive \& Skillman (2004)
who argue for a wider range of possible values of $Y_{\rm p}$,
indicated by the error bar on the right-hand side of the Figure.
The narrow box in the bottom panel shows the location of the
`Spite plateau', the narrow range of $^7$Li abundances measured
in old, metal-poor stars in the Galactic halo by Ryan et al. (2000),
and which is also well below the expected value of ($^7$Li/H)$_0$.
}
\label{}
\end{figure}

The situation is even worse for $^7$Li whose abundance in the most
metal-poor stars of the Galactic halo exhibits a narrow range 
(the well-known `Spite plateau') which is approximately three times
lower than the primordial value expected for $\Omega_{\rm b} = 0.044$
(Ryan et al. 2000). A number of possible explanations have been 
considered, including uncertainties in the effective temperatures
of the stars (e.g. Bonifacio et al. 2002), 
Li depletion in their atmospheres through mixing
with material from the stellar interior (e.g. Pinsonneault et al. 2002),
and errors in the relevant nuclear reaction rates (Coc et al. 2004).
None of these, however, is fully convincing.
Naively one would expect the first two effects to result in 
a dispersion of the Li abundances in metal-poor stars---in 
contrast with the narrow range observed, while the third would
require the rate for the $^7$Be($d, p$)2~$^4$He reaction 
(which competes with $^7$Be($n, p$)$^7$Li for the destruction of $^7$Be)
to be revised upwards by more than two orders of magnitude.

In conclusion, if D$_0$, the CMB power spectrum, and the Ly$\alpha$
forest average transmission give us the correct value of
$\Omega_{\rm b}$---as  seems likely given the concordance of these 
three very different methods---we are left with an internal tension 
between the abundances of the light elements created in BBNS. 
For $^4$He we can appeal to systematic uncertainties in the 
measurements, but $^7$Li remains a puzzle.

\section{Looking Ahead}
The convergence of different methods upon the same value of 
$\Omega_{\rm b}$ naturally raises the question as to whether it is worthwhile
searching for additional QSO absorption systems where the 
primordial abundance of D may be measured, especially
given the considerable investment in telescope time required.
In my view this is still an important goal, as I now explain.

%
% FIGURE 6
%
\begin{figure}[!ht]
\vspace*{-0.5cm}
\hspace{1.5cm}
\plotfiddle{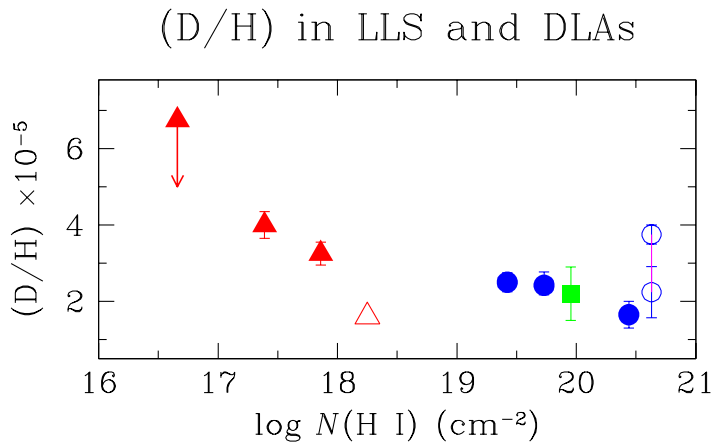}{5.5cm}{270}{60}{60}{-225}{250}
\vspace{0.35cm}
\caption{D/H measurements in QSO absorbers plotted against neutral hydrogen column
density. The symbols have the same meaning as in Figure~2. There is a hint that
D/H may be lower in DLAs (filled circles) than in LLS (filled triangles), but more
D detections are required to establish whether the trend is real or merely 
a statistical coincidence.
}
\label{}
\end{figure}

The first point is that the dispersion among the five
measurements of D/H which are considered most 
reliable in Figure 3 is higher than expected 
from their errors. This is not
an unusual situation in astronomy, and the natural conclusion
would be that the errors on the individual measurements
have been underestimated, were it not for the fact that
measures of D/H in the Galactic interstellar medium also
show a dispersion which, as discussed above, is now thought to be real. 
Neither of the explanations put forward for the local
dispersion ---dust depletion and chemical inhomogeneities---are
likely to apply to the high redshift absorption systems which
are significantly metal- and dust-poor (e.g. Pettini et al. 1997a,b).
Nevertheless, until the source(s) of variation---at both high
and low redshift---are identified, we cannot be totally confident 
of our measurement of D$_0$.

As an illustration of this, the data in Figure 2 are replotted in Figure 6
as a function of neutral hydrogen column density $N$(\ion{H}{1})
to show that, among the five measurements considered to be most reliable,
there is a hint of a trend with $N$(\ion{H}{1}), 
in the sense that D/H is lower in DLAs than LLS. 
This trend could be totally spurious, and a single
future measurement could show it to have been simply 
an artifact of the small number statistics.
On the other hand, one can certainly think of reasons why LLS may give
systematically higher estimates of D/H. The chances of contamination 
with hydrogen interlopers increase roughly as 1/$N$(\ion{H}{1})$^{0.5}$
(given that the number density of Ly$\alpha$ forest lines per unit column density
interval scales as $N$(\ion{H}{1})$^{-1.5}$, and are therefore lower
for DLAs with $N$(\ion{H}{1})\,$ \geq 2 \times 10^{20}$\,cm$^{-2}$
than for LLS with $N$(\ion{H}{1})\,$ \geq 3 \times 10^{17}$\,cm$^{-2}$.
Furthermore, the measurement of $N$(\ion{H}{1}) is more straightforward
in DLAs, where it relies on fitting the profile of the damping wings
of the Ly$\alpha$ absorption line, than in optically thick LLS where one
has to fit simultaneously a large number of saturated Lyman series lines.
Possibly, then, the values of D/H in DLAs should be given a higher
weighting than those in LLS in order to obtain the best estimate of the
primordial D abundance. I am not advocating such a course of action here,
but simply pointing out that additional measurements of D/H in QSO
absorption line systems at high $z$ are required to exclude (or confirm)
this possibility.

%
% FIGURE 7
%
\begin{figure}[!ht]
%\vspace*{-0.5cm}
%\hspace{1.25cm}
%\includegraphics[width=0.6\textwidth,angle=270]{mp_fig7.eps}
\plotfiddle{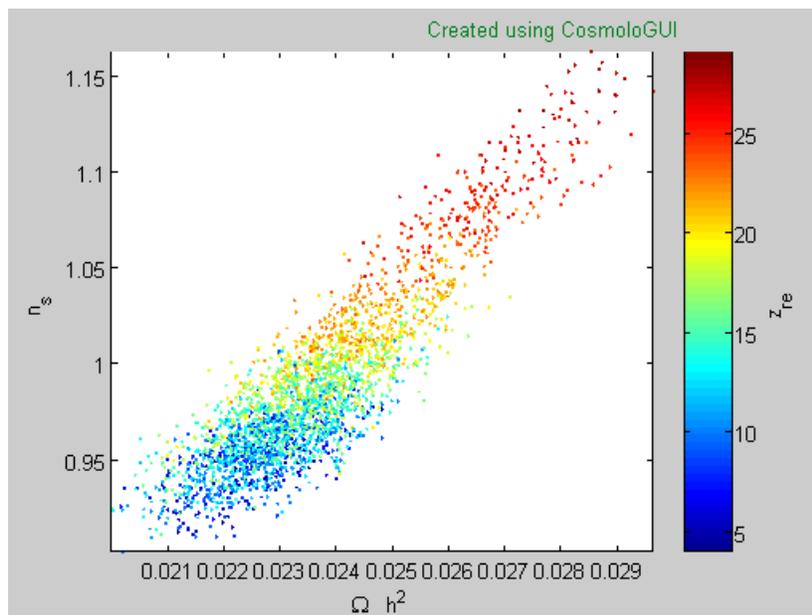}{7.5cm}{270}{60}{60}{-248}{290}
\vspace{0.15cm}
\caption{Interdependence of $\Omega_{\rm b}$ on other cosmological
parameters---in this case the power-law index of primordial fluctuations $n_s$
and the redshift of reionisation $z_{\rm re}$. All the combinations considered here
give acceptable fits to the \emph{WMAP} temperature (TT) and temperature-polarization
(TE) angular power spectra. (Reproduced from http://www.ast.cam.ac.uk/$\sim$sarah/cosmologui).
}
\label{}
\end{figure}

A second, and perhaps more fundamental, point is that among the various
avenues which lead to the determination of $\Omega_{\rm b}$, the primordial abundance
of deuterium is the most straightforward one, since it measures a combination
of only two cosmological parameters, $\Omega_{\rm b}$ and $H_0^2$. 
In contrast, other methods generally measure combinations of several such parameters.
In Figure 3, Gary Steigman showed the dependence upon $\Omega_{\rm b}$  of 
the amplitudes (and locations) of the acoustic peaks of the 
CMB angular power spectrum {\em by keeping fixed all other
relevant parameters at their most likely values.\/} 
In contrast, Figure 7 
shows the interdependence of the baryon density $\Omega_{\rm b}$, 
the power-law index of primordial
fluctuations $n_s$, and the optical depth to Thomson scattering
which determines the redshift of reionisation. Now we see that the 
CMB value of $\Omega_{\rm b} = (0.045 \pm 0.002)$ corresponds to
the `best' solutions for $n_s = 0.93$ (a nearly scale-invariant spectrum of 
fluctuations) and $z_{\rm re} = 17$ (see Spergel et al. 2003), 
but that other combinations of these parameters are in fact admitted. 
Similarly, the Ly$\alpha$ forest optical depth method of Tytler et al. (2004)
determines simultaneously $\Omega_{\rm b}$, $\sigma_8$
(the rms mass density fluctuation averaged over $8 h^{-1}$\,Mpc spheres),
and the ionisation rate of hydrogen at a given redshift.
%one then finds the combination of the three parameters which best fits the observations.
It follows, then, that if we knew $\Omega_{\rm b}$ with confidence
from the primordial abundance of deuterium, we would be able to narrow down
the allowed parameter space for other important cosmological quantities,
and this is in itself strong motivation for improving the still very limited
statistics on D/H at high $z$.
The Sloan Digital Sky Survey will, when completed, more than double the 
known number of DLAs (e.g. Prochaska \& Herbert-Fort 2004). 
I expect that, with perseverance,
it will be possible to identify in that treasure-trove of spectra 
several new absorption systems suitable for the determination of D$_0$.

\section{Summary}
We have come a long way since that pioneering measurement by Rogerson \& York
of the interstellar abundance of deuterium with 
{\em Copernicus\/} more than thirty years ago.
The number of such measurements now approaches fifty,
thanks in particular to the capabilities of {\em FUSE\/}
and the GHRS and STIS instruments on the {\em Hubble Space Telescope\/}.
With large ground-based telescopes we have been able to probe
high redshift clouds where D is still close to its primordial
abundance. New methods have been exploited to determine the 
density of baryons, the most impressive of which is the mapping
of the temperature anisotropies in the cosmic microwave background
over a wide range of angular scales. 
Bringing all of these developments together we find that many
aspects of the overall picture fit together remarkably well,
giving us confidence in the validity of the whole cosmological
framework. Others still provide challenging puzzles, particularly
the unexplained dispersion in the local determinations 
of D/H and the very low abundance of $^7$Li in some of the oldest stars
of our Galaxy. But I am optimistic that we will not have to wait another
three decades to iron out these remaining wrinkles 
in our understanding 
of the abundances of the light elements.

\acknowledgements I am indebted to Richard Sword and Mark Wilkinson for their
generous help with the preparation of the figures. 
I acknowledge useful conversations with Sarah Bridle and Bernard Pagel.
Finally, I should like to express 
my gratitude to the organisers of this most enjoyable meeting for including me 
among the participants.

% If you wish to use BiBTeX uncomment and fill in the .bib file name.  Note that 
% we are still (July 23) waiting for input from the ASP as to which 
% "bibliographystyle" to use.  "natbib" is unlikely to be the right one, but is 
% left here as a place holder.

%\bibliography{bib-file}
%\bibliographystyle{natbib}

% For using the "thebibliography" environment use these.
% See the "Authors Instructions" for details.

\end{document}